\documentclass[10pt,aps,prd,twocolumn]{revtex4-1}
\usepackage{acronym,amsmath,graphicx,textcomp}

\begin{document}

\title{How to search for gravitational waves from $r$-modes of known pulsars}
\author{Santiago Caride}
\affiliation{
Department of Physics and Astronomy,
Texas Tech University,
Lubbock, Texas, 79409-1051, USA
}
\author{Ra Inta}
\altaffiliation[Current address: ]{
Accelebrate,
925B Peachtree Street, NE,
PMB 378,
Atlanta, GA 30309-3918,
USA
}
\affiliation{
Department of Physics and Astronomy,
Texas Tech University,
Lubbock, Texas, 79409-1051, USA
}
\author{Benjamin J. Owen}
\affiliation{
Department of Physics and Astronomy,
Texas Tech University,
Lubbock, Texas, 79409-1051, USA
}
\author{Binod Rajbhandari}
\affiliation{
Department of Physics and Astronomy,
Texas Tech University,
Lubbock, Texas, 79409-1051, USA
}

\begin{abstract}
Searches for continuous gravitational waves from known pulsars so far have
been targeted at or near the spin frequency or double the spin frequency of
each pulsar, appropriate for mass quadrupole emission.
But some neutron stars might radiate strongly through current quadrupoles via
$r$-modes, which oscillate at about four thirds the spin frequency.
We show for the first time how to construct searches over appropriate ranges
of frequencies and spin-down parameters to target $r$-modes from known
pulsars.
We estimate computational costs and sensitivities of realistic $r$-mode
searches using the coherent $\mathcal{F}$-statistic, and find that feasible
searches for known pulsars can beat spin-down limits on gravitational wave
emission even with existing LIGO and Virgo data.
\end{abstract}

\maketitle

\acrodef{EM}{electromagnetic wave}
\acrodef{GW}{gravitational wave}
\acrodef{O1}{first observing run}
\acrodef{O2}{second observing run}
\acrodef{PSD}{power spectral density}
\acrodef{SFT}{short Fourier transform}

\section{Introduction}

In terms of strain, the most sensitive searches for \acp{GW} are those for
continuous waves, signals emitted by spinning neutron stars which need not be
in binaries.
The most sensitive searches for continuous \acp{GW} are those for known
pulsars, for which a timing solution derived from \ac{EM} observations
allows coherent integration of years of data~\cite[and references
therein]{Riles:2017evm}.
Most known pulsar searches (most recently~\cite{Authors:2019ztc}) have
targeted \ac{GW} frequencies precisely double (or occasionally equal to) the
observed spin frequency of each pulsar, based on electromagnetic pulse timing
and assuming an emission model of a ``mountain''---a mass quadrupole rotating
with the star.
Some known pulsar searches (most recently~\cite{Abbott:2019bed}) have instead
targeted narrow frequency bands of a fraction of a~Hz, losing some sensitivity
and costing more than fully targeted search, but allowing for some
uncertainties in the \ac{EM} timing parameters and in the physics such as the
possibility of free precession.

Neutron stars might also emit \acp{GW} via $r$-modes, rotation-dominated
quasi-normal modes driven unstable by gravitational radiation with frequencies
roughly 4/3 the spin frequency of the star~\cite[and references
therein]{Paschalidis:2016vmz}.
There are many uncertainties in the damping mechanisms that compete with the
instability and in the amplitudes attainable by $r$-modes due to nonlinear
hydrodynamics and other saturation mechanisms.
Within those uncertainties, $r$-modes might be oscillating at relatively low
amplitudes in fast spinning young pulsars up to several thousand years after
their birth, in rapidly accreting neutron stars, and in millisecond pulsars
perhaps a long time after accretion stops~\cite[and references
therein]{Glampedakis:2017nqy}.

So far \ac{GW} searches have only set upper limits on $r$-modes in broad band
(hundreds to thousands of~Hz) searches for non-pulsing neutron stars (most
recently~\cite{Abbott:2018qee}), but searches could be done for $r$-modes from
known pulsars too.
The main issue is determining the frequency band to search.
The uncertainty in $r$-mode frequency for a known pulsar is typically a
few~Hz~\cite{Idrisy:2014qca}, broader than previous narrow band pulsar
searches but not as broad as previous $r$-mode searches.
For long integrations, which are the most sensitive searches, some thought
needs to be given to spin-down parameters as well.

The main caveat is that $r$-modes might not truly be unstable (damping might
beat driving) in all or most neutron stars once all damping mechanisms are
taken into account.
Another caveat is that $r$-modes might be unstable but saturate at amplitudes
too small to be detectable with present and near-future detectors.
Predictions of saturation amplitudes~\cite{Arras:2002dw} are indeed too small
to detect for most pulsars and present detectors~\cite{Owen:2010ng}.
But calculations of saturation amplitude might be wrong and \ac{GW} detectors
are improving.
Also, it takes some time to develop and refine a \ac{GW} search, so it is
worthwhile to start now.

In this article we describe how to perform searches for \acp{GW} from
$r$-modes of known pulsars using minimal adaptations of existing code.
In particular the directed search pipeline used most recently in
Ref.~\cite{Abbott:2018qee}, based on the implementation of the
$\mathcal{F}$-statistic in LALSuite~\cite{LALSuite}, is easily adaptable for
this purpose.
We present a method for choosing a search parameter space and estimate
computational costs and sensitivities using that method.
(The parameter space was partially estimated once before~\cite{Ian}.)
We show that interesting searches of data from LIGO's \ac{O1} and \ac{O2} are
feasible already, and that future searches of more sensitive data sets will be
even more interesting.
Along the way we highlight the main issues affecting realistic observations,
which leads naturally to a list of suggestions for future work both by
theorists and by data analysts.

\section{Assumptions}

\subsection{Physics}

We assume that the \ac{GW} frequency evolution $f(t)$ in the reference frame
of the solar system barycenter is
\begin{equation}
\label{ft}
f(t) = f\left( t_0 \right) + \dot f\left( t_0 \right) \left( t-t_0 \right) +
\frac{1}{2} \ddot f\left( t_0 \right) \left( t-t_0 \right)^2,
\end{equation}
where $t_0$ is some reference time (often the beginning of the observation),
dots indicate time derivatives, and we shall use a simple $f$ to indicate
$f(t_0)$ from now on.
Physically, Eq.~(\ref{ft}) assumes that the signal frequency does not change
too fast (such as from glitches or higher derivatives) or too erratically
(such as from timing noise).
Timing noise is unlikely to be an issue for the integration times (a year or
less) considered here~\cite{Ashton:2014qya}.
Glitches can be avoided by checking \ac{EM} observations.
As we shall see below, higher derivatives are not a problem, and for some
searches even the second derivative is not needed.

Throughout we consider the lowest order (current quadrupole) $r$-mode, since
it is the fastest driven by \ac{GW} emission and the least damped by most
forms of viscosity.

We shall make frequent use of the ``spin-down limits''~\cite{Owen:2010ng} on
intrinsic \ac{GW} strain~\cite{Jaranowski:1998qm}
\begin{equation}
h_0^\mathrm{sd} \simeq 1.6\times10^{-24} \left( \frac{\mbox{1 kpc}}{r} \right)
\left( \frac{\left|\dot f\right|} {10^{-10}\mbox{ Hz/s}} \frac{\mbox{100 Hz}}
{f} \right)^{1/2}
\label{h0sd}
\end{equation}
(where $r$ is the distance to the pulsar) and on the $r$-mode amplitude
parameter~\cite{LMoO:1998prl}
\begin{equation}
\alpha_\mathrm{sd} \simeq 0.033 \left( \frac{\mbox{100 Hz}} {f} \right)^{7/2}
\left( \frac{|\dot{f}|} {10^{-10}\mbox{ Hz\,s}^{-1}} \right)^{1/2}.
\end{equation}
These correspond to the assumption that all of the observed $\dot f$ is due to
\ac{GW} emission via $r$-modes.
That is clearly an unrealistic assumption, especially when considering the
observed values of second derivatives too; but these limits serve as useful
milestones for search sensitivity.
These numerical forms of the limits assume certain neutron star structure
parameters and also assume that the ratio of \ac{GW} frequency to spin
frequency is 4/3, so they are uncertain by a factor of two or
so~\cite{Owen:2010ng}.
While they should not be taken too literally, these spin-down limits give a
rough idea of which \ac{GW} searches are most interesting.

There are many debates on the growth and damping timescales of the $r$-modes,
and on the saturation amplitude (which determines the long term strength of
\ac{GW} emission).
We largely bypass them, though we note that $\alpha$ is predicted to saturate
at order $10^{-4}$ or lower~\cite{Arras:2002dw}.
Theoretical uncertainties in these quantities are great and might best be
resolved by observations which do not rely too much on theoretical guidance to
pick targets.
Attempts at relatively model independent predictions or connections to
electromagnetic observations sometimes favor or disfavor one pulsar or
another.
Alford and Schwenzer~\cite{Alford:2012yn} argue that $r$-modes in young
neutron stars spinning down shut off above 60~Hz under a wide variety of
conditions, making PSR~J0537\textminus6910 the best candidate.
They also propose that the braking index (see below) could be as low as four
in some $r$-mode dominated pulsars rather than seven as implied for
constant~$\alpha$ evolution~\cite{Owen:1998xg}.
Certainly J0537\textminus6910 has the best (lowest)
$\alpha_\mathrm{sd}$ for all young pulsars~\cite{Owen:2010ng} (order
$10^{-1}$), and there are hints that the braking index between its frequent
glitches might be seven~\cite{Andersson:2017fow}.
There are also arguments that millisecond pulsars will emit \acp{GW} from
$r$-modes for a long time, but only at small amplitudes even if they have high
spin-down limits~\cite{Bondarescu:2013xwa, Alford:2014pxa}.
And temperature observations of some pulsars might indicate small $r$-mode
amplitudes due to constraints on viscous heating~\cite{Schwenzer:2016tkf}.
But as observers we note that the universe holds many surprises, and we
consider searches for any pulsar with an attainable spin-down limit.

An $r$-mode emits \acp{GW} at the mode's oscillation frequency $f$ in an
inertial frame of reference.
This frequency is a function of star's spin frequency $\nu$ of the
form~\cite[e.g.]{Yoshida:2004gk}
\begin{equation}
\label{fnu}
f/\nu = A - B \left( \nu / \nu_K \right)^2 + O(\nu)^4,
\end{equation}
where we have chosen the signs so that $A>0$ and $B>0.$
(The sign of the second term is generic to retrograde modes, such as all
$r$-modes, and is due to the effects of gravitational redshift and dragging of
inertial frames on the Coriolis force~\cite{Paschalidis:2016vmz}.)
Both parameters depend on the (usually unknown) mass $M$ and radius $R$ of the
neutron star and the still uncertain equation of state.
The rapid rotation calculation of $r$-mode frequencies in
Ref.~\cite{Yoshida:2004gk} indicates that the $O(\nu)^4$ remainder in
Eq.~(\ref{fnu}) is negligible except (for some stars) when the star is
spinning almost at its Kepler frequency $\nu_K,$ the frequency at which
centrifugal force tears it apart.
Most pulsars, including those we find most interesting for this analysis, spin
at much less than the 716~Hz of the fastest observed
frequency~\cite{Manchester:2004bp}, so we will neglect the $O(\nu)^4$
remainder in Eq.~(\ref{fnu}).
We also assume that any backbending (nonmonotonic $f$ vs.\ $\nu$) due to a
possible phase transition~\cite{Glendenning:1997fy} occurs at much higher
frequencies and does not appreciably affect Eq.~(\ref{fnu}).

In Eq.~(\ref{fnu}) we neglect many physical effects which should have only a
small effect on the ranges of $A$ and $B.$
We assume that any differential rotation has a negligible effect on the mode
frequency.
In practice this seems likely to be true, even though the $r$-mode tends to
generate a small amount of differential rotation analogous to Stokes
drift~\cite{Friedman:2017wfi}.
We assume that the magnetic field's direct effect (through restoring force) on
the $r$-mode frequency is small.
Several studies (most recently~\cite{Jasiulek:2016epr}) indicate that this is
true for the relatively low magnetic fields of pulsars spinning in the LIGO
band.
While superfluidity generally has only a tiny effect on the mode
frequencies~\cite{Lindblom:1999wi}, it does split each normal fluid $r$-mode
into two where the neutron and proton fluids are co-moving or
counter-moving~\cite{Andersson:2001bz}; and the latter mode has an additional
restoring force due to entrainment of the two fluids.
These modes in general have slightly different frequencies, but the
counter-moving modes tend to be much more damped by mutual friction.
Thus we can assume that the \ac{GW} emission is almost all through co-moving
modes, which have frequencies almost identical to normal fluid modes.

More serious is the issue of avoided crossings between $r$-modes and other
modes.
For example, Levin and Ushomirsky~\cite{Levin:2000vq} pointed out that, as a
star spins down, the $r$-mode and a torsional mode of the solid crust
(vibrating at of order 100~Hz in a nonrotating star) will swap identities.
More sophisticated calculations~\cite[e.g.]{Glampedakis:2006ap} support this
idea.
A similar issue arises with the coupling between the $r$-modes and the buoyant
force responsible for $g$-modes~\cite{Kantor:2017xuo}.
In either case, in the vicinity of the other mode's frequency the avoided
crossing introduces large errors into the simple $r$-mode frequency dependence
posited in Eq.~(\ref{fnu}).
We shall neglect this, and assume that the pulsars we consider have $\nu$
safely away from the major avoided crossings.
In a search for many pulsars, most of them are likely to satisfy this
assumption; but it is a potentially serious issue worthy of more research.

The other potentially major effect is the coherence time of the $r$-modes.
If mode-mode coupling calculations such as Ref.~\cite{Brink:2004kt} are
correct, a saturated $r$-mode exists in rough equilibrium with two ``daughter
modes'' but may occasionally undergo abrupt phase shifts~\cite{Ira}.
A similar situation could hold if one attempts to take advantage of the
relatively broad band of frequencies to search for $r$-modes from a pulsar
which does not have \ac{EM} timing contemporary with a \ac{GW} data run---the
pulsar could have glitched during the \ac{GW} run, introducing a phase error
at a random time.
Such phase errors would reduce the sensitivity of coherent data analysis
methods described below~\cite{Ashton:2017wui}.

\subsection{Data analysis}

We assume a search method based on coherent integration using a minimal
adaptation of the code used in several broad band directed searches, most
recently Ref.~\cite{Abbott:2018qee}.
This code implements the multi-interferometer
$\mathcal{F}$-statistic~\cite{Jaranowski:1998qm, Cutler:2005hc}, which
combines matched filters in such a way as to account for the daily changes of
the interferometers' beam patterns as the Earth rotates.
The $\mathcal{F}$-statistic also quickly maximizes signal-to-noise ratio over
the (typically unknown) angles describing the orientation of the neutron
star's spin axis.
In Gaussian noise, $2\mathcal{F}$ is drawn from a $\chi^2$ distribution with
four degrees of freedom.
In the presence of a signal, the $\chi^2$ is noncentral and the power
signal-to-noise ratio (if large) is approximately $\mathcal{F}/2.$

For some pulsars a wind nebula indicates the orientation of the star's spin
axis, and in that case a similar statistic called the
$\mathcal{G}$-statistic~\cite{Jaranowski:2010rn} can achieve slightly better
sensitivity.
Since the $\mathcal{G}$-statistic is not included in the implementation of the
$\mathcal{F}$-statistic in LALSuite~\cite{LALSuite} used in
Ref.~\cite{Abbott:2018qee}, we do not consider it further here; but it is a
natural avenue of future improvement.

We assume that the \ac{GW} search can use a single sky position.
Pulsar positions are typically known to sub-arcsecond
precision~\cite{Manchester:2004bp} and the sky resolution of a directed
continuous \ac{GW} search is two orders of magnitude less precise at the
frequencies of most pulsars~\cite[e.g.]{Abbott:2018qee}.

We do not assume we have a coherent \ac{EM} pulsar timing solution throughout
the \ac{GW} observation.
Narrow band searches for pulsars such as~\cite{Abbott:2019bed} are already
broad enough to extrapolate \ac{EM} timing from old observations, and since
our search bands turn out to be broader they are even more robust.
The main worry is whether there is a glitch during the \ac{GW} observation, in
which case it will effectively cut the integration time and reduce the
signal-to-noise ratio~\cite{Ashton:2017wui}.

We do assume that, in cases where the pulsar is a component of a binary, the
binary's orbital parameters are known well enough to avoid requiring a search
over them.
This is true for most binary pulsars except those in low mass x-ray binaries.
Since those are accreting and therefore the spin can undergo random walks, we
neglect them for our present purposes.

\section{Parameter space}

Under the assumptions stated above, the parameter space of a search consists
of ranges of $f,$ $\dot f,$ and possibly $\ddot f.$
With some uncertainties, these can be calculated as functions of the
\ac{EM}-observed spin $\nu$ and spin-down parameters $\dot\nu$ and $\ddot\nu.$

\subsection{General expressions}

Assuming that $A$ and $B$ do not change appreciably with time, the time
derivatives of Eq.~(\ref{fnu}) yield
\begin{eqnarray}
\label{fdot}
\dot f / \dot \nu &=& A - 3B \left( \nu/\nu_K \right)^2,
\\
\label{fddot}
\ddot f / \ddot \nu &=& A - \left( 3 + 6/n \right) B \left( \nu/\nu_K
\right)^2,
\end{eqnarray}
where $n = \nu \ddot \nu / \dot \nu^2$ is the braking index of the pulsar.
Thus an observation of $(\nu, \dot\nu, \ddot\nu)$ and calculations of $\nu_K$
and the ranges of $(A,B)$ determine the ranges of $(f, \dot f, \ddot f)$ to be
searched---in principle.
In practice $\ddot\nu$ tends to have significant uncertainties, affecting the
choice of $\ddot f$ as discussed below.

To get the frequency range of a search we insert the ranges of $A$ and $B$
into Eq.~(\ref{fnu}) and---for now---assume that $\nu_K$ is known to obtain
\begin{eqnarray}
(f/\nu)_{\min} &=& A_{\min} - B_{\max} \left( \nu/\nu_K \right)^2,
\\
(f/\nu)_{\max} &=& A_{\max} - B_{\min} \left( \nu/\nu_K \right)^2.
\end{eqnarray}

To determine the range of $\dot{f}$ for a given $f,$ note that
Eqs.~(\ref{fnu}) and~(\ref{fdot}) can be combined to write
\begin{equation}
\dot f / \dot \nu = f / \nu - 2B \left( \nu/\nu_K \right)^2.
\end{equation}
Then we simply have the range
\begin{eqnarray}
\left( \dot{f} / \dot\nu \right)_{\min} &=& f / \nu - 2B_{\max} \left(
\nu/\nu_K \right)^2,
\\
\left( \dot{f} / \dot\nu \right)_{\max} &=& f / \nu - 2B_{\min} \left(
\nu/\nu_K \right)^2.
\end{eqnarray}

Note that Eqs.~(\ref{fnu}) and~(\ref{fdot}) combined determine $A$ and
$B/\nu_K^2$ as
\begin{eqnarray}
A &=& \left( 3f / \nu - \dot{f} / \dot{\nu} \right) /2,
\\
B/\nu_K^2 &=& \left( f / \nu - \dot{f} / \dot{\nu} \right) / \left( 2\nu^2
\right).
\end{eqnarray}
These relations can be used for parameter estimation from a \ac{GW} detection:
Once $f$ and $\dot f$ are known, we find $A$ and $B/\nu_K^2,$ which in turn
can yield information on $M$ and $R$ and the equation of
state~\cite{Yoshida:2004gk, Idrisy:2014qca}.
The equations for $A$ and $B$ also can be used to write
\begin{equation}
\ddot{f} / \ddot{\nu} = \dot{f} / \dot{\nu} - (3/n) \left( f / \nu - \dot{f} /
\dot{\nu} \right),
\end{equation}
which in principle uniquely determines $\ddot f$ in terms of $f,$ $\dot f,$
and \ac{EM}-observed quantities.

In practice, $\ddot\nu$ (or equivalently $n$) measurements are available only
for a few pulsars; and even then they may have large errors when measured over
short baselines.
For example, the monthly fits to $\ddot\nu$ for the Crab pulsar provided by
Jodrell Bank can vary by a factor of a few from the long term average and can
even change sign~\cite{Crab2015}.
It is not clear how much of this timing noise is due to magnetospheric effects
and how much is due to a genuine fluctuating torque on the star.
Since the $r$-mode frequency is determined mainly by the Coriolis force, we
are interested in the latter but not the former.
However at the moment we wish to be cautious in our choice of parameter space.
As a practical matter, current codes including that used in
Ref.~\cite{Abbott:2018qee} cut the parameter space into computing batch jobs
in a way such that the range of $\ddot f$ can depend on $f$ but not on $\dot
f.$
Plugging in the full range of $\dot f$ we can get
\begin{eqnarray}
\left( \ddot f / \ddot\nu \right)_{\min} &=& f/\nu - 2(1+3/n) B_{\max} \left(
\nu/\nu_K \right)^2,
\\
\left( \ddot f / \ddot\nu \right)_{\max} &=& f/\nu - 2(1+3/n) B_{\min} \left(
\nu/\nu_K \right)^2
\end{eqnarray}
as functions of $f.$
To err on the safe side by covering more parameter space, we can take the
minimum $\ddot f$ as zero.
Since our ``safe side'' $B_{\min}$ vanishes (see below), the maximum $\ddot f$
can be taken to be simply $\ddot\nu f/\nu,$ using the highest $\ddot\nu$
observed during the \ac{GW} integration.

At the moment these overly broad parameter ranges are not a concern, because
\ac{O1} searches are computationally cheap and even \ac{O2} searches are not
extravagant (see below).
These parameter ranges could be refined later for longer searches, when
computational cost is more of an issue, for example by calculating the range
of $B$ and exploring its consequences.

\subsection{Numerical ranges of parameters}

The range of $A$ is fairly well known.
The most recent calculation~\cite{Idrisy:2014qca} used the general
relativistic slow rotation approximation~\cite{Lockitch:2000aa,
Lockitch:2002sy} to compute $A$ for a variety of neutron-star equations of
state, obtaining 1.39 $\le A \le$ 1.57 depending almost purely on $M/R.$
Since that calculation was published, the big new constraint on the neutron
star equation of state is the lack of a large tidal effect in the binary
neutron-star merger GW170817~\cite{TheLIGOScientific:2017qsa}.
This disfavors large radii and low $A.$
But for the rest of this paper, to be conservative (cover a wide range of
parameters), we shall use the $A_{\min}$ and $A_{\max}$ quoted above.

The range of $B$ is less well known than the range of $A.$
The best general relativistic calculation~\cite{Yoshida:2004gk} drops the slow
rotation approximation, but adds the Cowling approximation (neglecting the
metric perturbation) and gives numbers only for two equations of state and two
$M/R$ values.
And the equations of state are polytropes rather than realistic equations of
state with the adiabatic index varying depending on the density.
The errors due to the Cowling approximation can be estimated as a few percent,
which is not of too much concern here; but the uncertainty from the stellar
models is more serious.

We estimate the range of $B$ from Ref.~\cite{Yoshida:2004gk} as follows:
Their Eq.~(15) gives $B\left( \nu/\nu_K \right)^2$ as 1.23--1.95 times the
ratio of kinetic to potential energy for the four stellar models considered.
For our purposes the most interesting model is their model $c,$ a polytrope of
adiabatic index~2 and $M/R=0.1$ which yields the number 1.95 (and for which
the slow rotation approximation is very accurate all the way to the Kepler
frequency).
In Fig.~2 of Ref.~\cite{Yoshida:2004gk} the sequence for model $c$ terminates
at a kinetic-to-potential energy ratio of 0.1, and this termination point
corresponds to $\nu=\nu_K,$ the ``Kepler frequency'' or maximum spin frequency
of the star.

Hence we can write $B \simeq 0.195$ for this stellar model, which should set a
safe upper limit on $B_{\max}$ for the following reasons:
The results of Ref.~\cite{Yoshida:2004gk} show that $B$ increases for smaller
$M/R$ and for lower adiabatic index (higher polytropic index).
The value $M/R=0.1$ for their model $c$ is smaller than post-GW170817 bounds,
indicating we are safe there.
An adiabatic index of~2 also errs on the safe side, since piecewise polytropic
fits to realistic equations of state~\cite{Read:2008iy} yield higher indices.
The bound on $B_{\min}$ is less clear without detailed calculations, but
$B_{\min}=0$ is well beyond the range quoted and should be safe.

Last we consider $\nu_K.$
The fastest observed spin frequency for a neutron star is about
716~Hz~\cite[and references therein]{Paschalidis:2016vmz}.
The Kepler frequency is expected to scale roughly as its Newtonian dependence
$M^{1/2} R^{-3/2},$ even in general relativity~\cite[and references
therein]{Paschalidis:2016vmz}.
Neutron star radii are roughly constant for a given equation of state, while
reliable mass measurements range over almost a factor of
two~\cite[and references therein]{Ozel:2016oaf}.
To err on the safe side (high $B_{\max}$), we assume that the 716~Hz pulsar is
on the high end of the mass range and our pulsar is on the low end so that it
has a lower Kepler frequency.
Then we can safely take $\nu_K = \mbox{716 Hz}/\sqrt{2} \simeq 506$~Hz, erring
on the safe side by assuming a possible factor of two difference in mass.

To summarize, we recommend for the moment a broad parameter space with ranges
\begin{eqnarray}
\label{range0}
f_{\min} = \nu \left( A_{\min} - B_{\max} \frac{\nu^2} {\nu_K^2} \right),
&\quad&
f_{\max} = \nu\, A_{\max},
\\
\label{range1}
\dot f_{\min} = -\dot\nu \left( \frac{f}{\nu} - 2B_{\max} \frac{\nu^2}
{\nu_K^2} \right),
&\quad&
\dot f_{\max} = -\dot\nu \frac{f}{\nu},
\\
\label{range2}
\ddot f_{\min} = 0,
&\quad&
\ddot f_{\max} = \ddot\nu \frac{f}{\nu},
\end{eqnarray}
where $\ddot\nu$ is the maximum value consistent with \ac{EM} observations,
and the other parameters are $A_{\min} = 1.39,$ $A_{\max} = 1.57,$ $B_{\max} =
0.195,$ and $\nu_K = 506$~Hz.

\section{Computational cost}

We rely heavily on the search parameter space metric~\cite{Wette:2008hg}
\begin{equation}
g_{ij} = \pi^2 \left(
\begin{array}{ccc}
T^2/3 & T^3/6 & T^4/20 \\
T^3/6 & 4T^4/45 & T^5/36 \\
T^4/20 & T^5/36 & T^6/112
\end{array}
\right)
\end{equation}
where the indices are labeled in the order $(f, \dot f, \ddot f)$ and also can
be labeled 0, 1, 2.
Here $T$ is the time from beginning to end of the integration.
This metric controls the density and placement of templates (parameter values
of matched filters) by relating coordinate distances (parameter differences)
to loss of signal-to-noise ratio~\cite{Owen1996}.
The mismatch $g_{ij} \Delta \lambda^i \Delta \lambda^j$ between two signals
with parameters displaced by $\Delta \lambda^i$ is the fractional loss in
optimal power signal-to-noise ratio due to filtering one with the parameters
of the other.
In continuous \ac{GW} searches template banks are often constructed so that
the worst case mismatch between any signal and the nearest template is 0.2.
We calculate metric components neglecting the amplitude modulation of the
$\mathcal{F}$-statistic and including only phase terms, an approximation which
works well for integrations of many days~\cite{Prix:2006wm}.

To determine which pulsars require $\ddot f,$ we compute $g_{22} \ddot
f_{\max}^2$ to determine the maximum mismatch due to neglecting $\ddot f.$
(Note that this does not allow for the possible mitigating effect of varying
$f$ and $\dot f$ somewhat, and thus it is a conservative estimate.)
If the mismatch is comparable to or greater than 0.2, $\ddot f$ is needed.
First we evaluate this criterion using $\ddot\nu$ values taken from the ATNF
catalogue~\cite{Manchester:2004bp}.
In many cases these values are unknown or are known to be contaminated by
timing noise (such as when they are negative).
However the values of $\nu$ and $\dot\nu$ are typically well measured.
Hence we also check the need for $\ddot f$ using $\ddot\nu = 7 \dot\nu^2 /
\nu$ (implied by a braking index of 7), and if either this mismatch or the one
using the observed $\ddot\nu$ satisfies the criterion we consider a search
over $\ddot f$ to be necessary.

We consider three values of $T:$
First $1.12\times10^7$~s and $2.32\times10^7$~s, the lengths of \ac{O1} and
\ac{O2} respectively, then one year or $3.15\times10^7$~s which might be
characteristic of future LIGO runs~\cite{Aasi:2013wya}.
We only consider pulsars with $f_{\max}$ greater than 10~Hz since, even at
design sensitivity, LIGO noise increases rapidly below that frequency and the
$r$-mode amplitudes required to emit at the spin-down limit become enormous.
We find that for \ac{O1} the Crab, Vela, and several others need $\ddot f;$
while for \ac{O2} and a one-year integration most pulsars with measured
$\ddot\nu$ and several without it need $\ddot f.$

We test the need of the third frequency derivative using $g_{33} = \pi^2 T^8 /
2025$~\cite{Wette:2008hg}, the third derivative of $\nu$ from the ATNF
catalogue~\cite{Manchester:2004bp} when it is given, and the $n=7$ value of
$91 \dot\nu^3 / \nu^2$ for the third derivative when it is not given.
(The numerical factor 91 is $n(2n-1)$ in general, and can be obtained by
differentiating the definition of the braking index.)
For \ac{O1} no pulsar needs a third derivative.
For \ac{O2} no pulsar with a spin-down limit above the noise (see below) needs
a third derivative.
For a one year integration the Crab (alone of the pulsars detectable at the
spin-down limit) is on the edge of needing a third derivative.
Since this is a conservative estimate and the need can be mitigated by
slightly shortening $T$ and our focus here is on how to adapt current data
analysis codes, which do not include the third derivative, we do not address
third derivatives further here.

Under the assumption that two frequency derivatives are needed and three are
not, the proper volume of the parameter space $\sqrt{g} \int df\, d\dot f\,
d\ddot f$ integrated over the ranges in Eq.~(\ref{range0})--(\ref{range2}) is
approximately
\begin{equation}
\label{propvol}
\sqrt{g} \nu \left| \dot\nu \right| \ddot\nu B_{\max} \left( \nu/\nu_K
\right)^2 \left[ A_{\max}^2 - A_{\min}^2 \right].
\end{equation}
(Here we have dropped the $B$ terms in $f_{\min}$ and $f_{\max},$ since they
are small corrections, and $g$ indicates the determinant of the metric.)
To estimate the number of templates, we divide by the proper volume per
template~\cite{Owen1996},
\begin{equation}
V = \left( 2 \sqrt{\mu/3} \right)^3 \simeq 0.138
\end{equation}
for a three-dimensional template bank mismatch $\mu$ of 0.2.
In cases where $\ddot f$ is not needed, these expressions need to be modifed,
but those cases are so computationally cheap that they are not an issue.
In practice the number of templates is modified from these estimates by the
vagaries of the actual template placement code, typically dominated by the
problem of covering the edges of long narrow stretches of parameter space.
Our tests with LALSuite~\cite{LALSuite} show that a real search might use
three times as many templates as these ideal numbers.
Since this factor can vary for each search, we do not include it further or
attempt to estimate the numbers too precisely.

To get values for the number of templates, we use $\sqrt{g} \simeq T^6 \times
8.41\times10^{-3}$ where $T$ is measured in seconds.
We find that for \ac{O1} using ATNF values of $\ddot\nu,$ the Crab requires
$6\times10^9$ templates and the others generally require one or more orders of
magnitude fewer than the Crab.
Taking catalogue numbers at face value, the exception is J0537\textminus6910
which requires $1.5\times10^{10}$ templates.
Using the maximum $\ddot f$ derived from a braking index of 7, the Crab
triples to about $2\times10^{10}$ templates.
Using ATNF spin-downs, for \ac{O2} the Crab and PSR~J0537\textminus6910
require of order $5\times10^{11}$ and $1\times10^{12}$ templates, and for a
one year integration they require $3\times10^{12}$ and $7\times10^{12}.$
The latter number is the same as the first directed search for an isolated
neutron star~\cite{Abadie:2010hv}, whose cost was modest by the standards of
continuous \ac{GW} data analysis.

Again, these numbers are likely to be larger in reality due to the template
placement algorithms in LALSuite~\cite{LALSuite}.
And, although we make rough blanket statements here, for a real search each
pulsar needs some investigation into timing noise and glitches.
For example, while PSR~J0537\textminus6910 might be a very interesting pulsar
to search, it is known to glitch frequently and because it is visible only in
x-rays it is important to maintain satellite timing~\cite{Andersson:2017fow}.
Even with timing, if this pulsar glitches in the middle of a \ac{GW} observing
run, the run will need to be divided into segments.
For another example, since in \ac{O2} the noise performance of the Hanford
interferometer was usually significantly worse than the Livingston
interferometer at the frequencies of most pulsars with high spin-down limits,
some pulsar searches might use only data from Livingston with little loss in
sensitivity.

To convert template numbers to computational cost, we run a piece of the
latest directed search code used in~\cite{Abbott:2018qee} to estimate the
computational cost per \ac{SFT} of 30 minutes of data per template.
On the LIGO-Caltech cluster Broadwell and Skylake benchmarking nodes the cost
is generally somewhat less than 50~ns per \ac{SFT} per template, depending on
network activity and disk throughput.
For \ac{O1} the number of \acp{SFT} was about $6\times10^3.$
For \ac{O2} the number is about double, but we still use $7\times10^3$
assuming a search which does not integrate data from the Hanford
interferometer because its noise is significantly worse than that at
Livingston for the low frequencies considered here.
For a one year search of future data we assume two interferometers at a duty
cycle of 70\% each, comparable to the most stable past operation of the
interferometers, resulting in $2.5\times10^4$ \acp{SFT}.

Under these assumptions the cost of a Crab search is of order 500 core-hours,
$5\times10^4,$ or $1\times10^6$ for \ac{O1}, \ac{O2}, or one year
respectively.
This indicates that searching all pulsars for \ac{O1} and \ac{O2} is not a
computational problem, even though the number of templates is likely to be
larger in reality.
The one-year figure for the Crab is comparable to the total power used in a
bundle of recent directed searches~\cite{Abbott:2018qee}.
It is not outrageously expensive, but indicates that soon it will be desirable
to reduce the costs through a combination of theory and data analysis
innovations.

The density of templates per unit frequency is useful in estimating
sensitivity (below) and in load balancing the code.
Derived similarly to the proper volume~(\ref{propvol}) but omitting the $\int
df$ and dividing by the volume per template $V,$ this density is approximately
\begin{equation}
2\sqrt{g} B_{\max} \left( \nu/\nu_K \right)^2 \left| \dot\nu \right| \ddot\nu
f/\nu / V.
\end{equation}
For \ac{O1} at the maximum end of the frequency ranges, this takes the values of
$1.1\times10^9$ and $1.4\times10^9$\,Hz$^{-1}$ for the Crab and J0537
respectively.
For \ac{O2} the corresponding numbers are about $9\times10^{10}$ and
$1.1\times10^{11},$ and for a one year integration they are about
$5\times10^{11}$ and $7\times10^{11}.$

\section{Sensitivity}

\begin{figure*}
\includegraphics[angle=270,width=\linewidth]{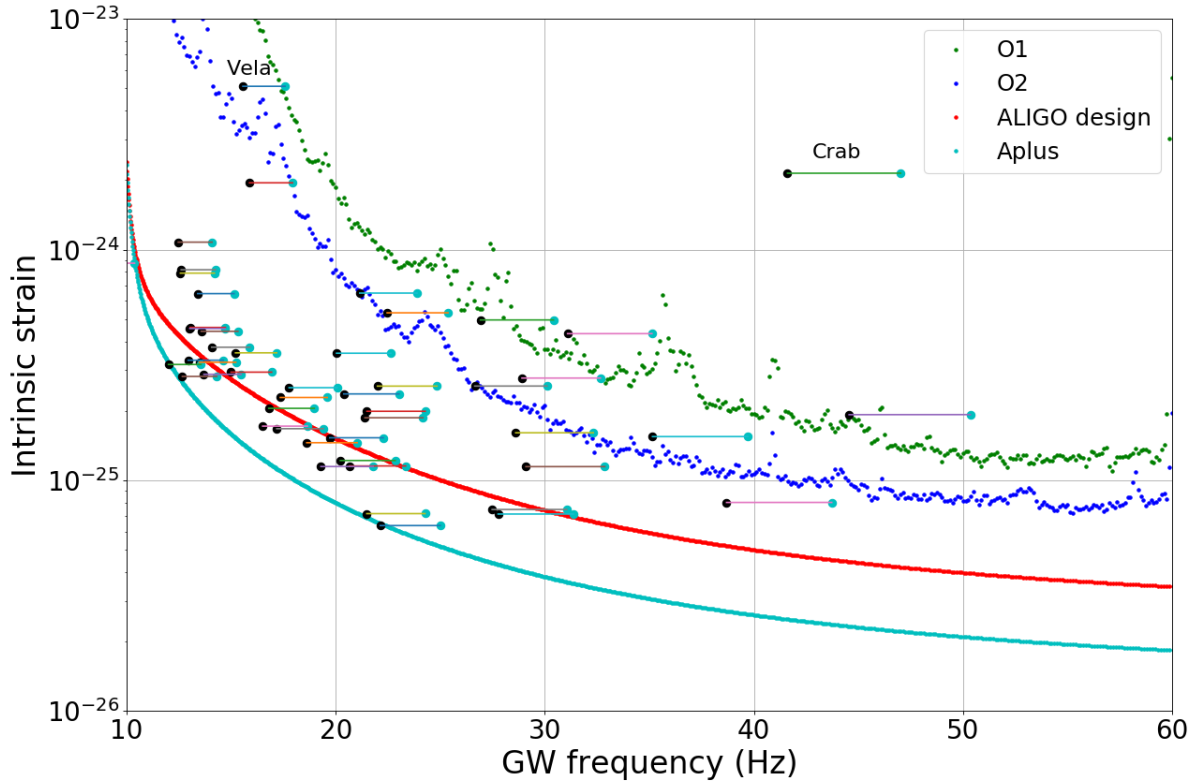}
\caption{
\label{fig1}
Spin-down limits for interesting pulsars (horizontal lines) and sensitivity
estimates (other curves), both in terms of intrinsic strain vs.\ \ac{GW}
frequency.
Spin-down limits are taken from Eq.~(\ref{h0sd}) in the text.
Sensitivity estimates are taken from Eq.~(\ref{h0Td}) and the paragraph
containing it.
}
\end{figure*}

\begin{figure*}
\includegraphics[angle=270,width=\linewidth]{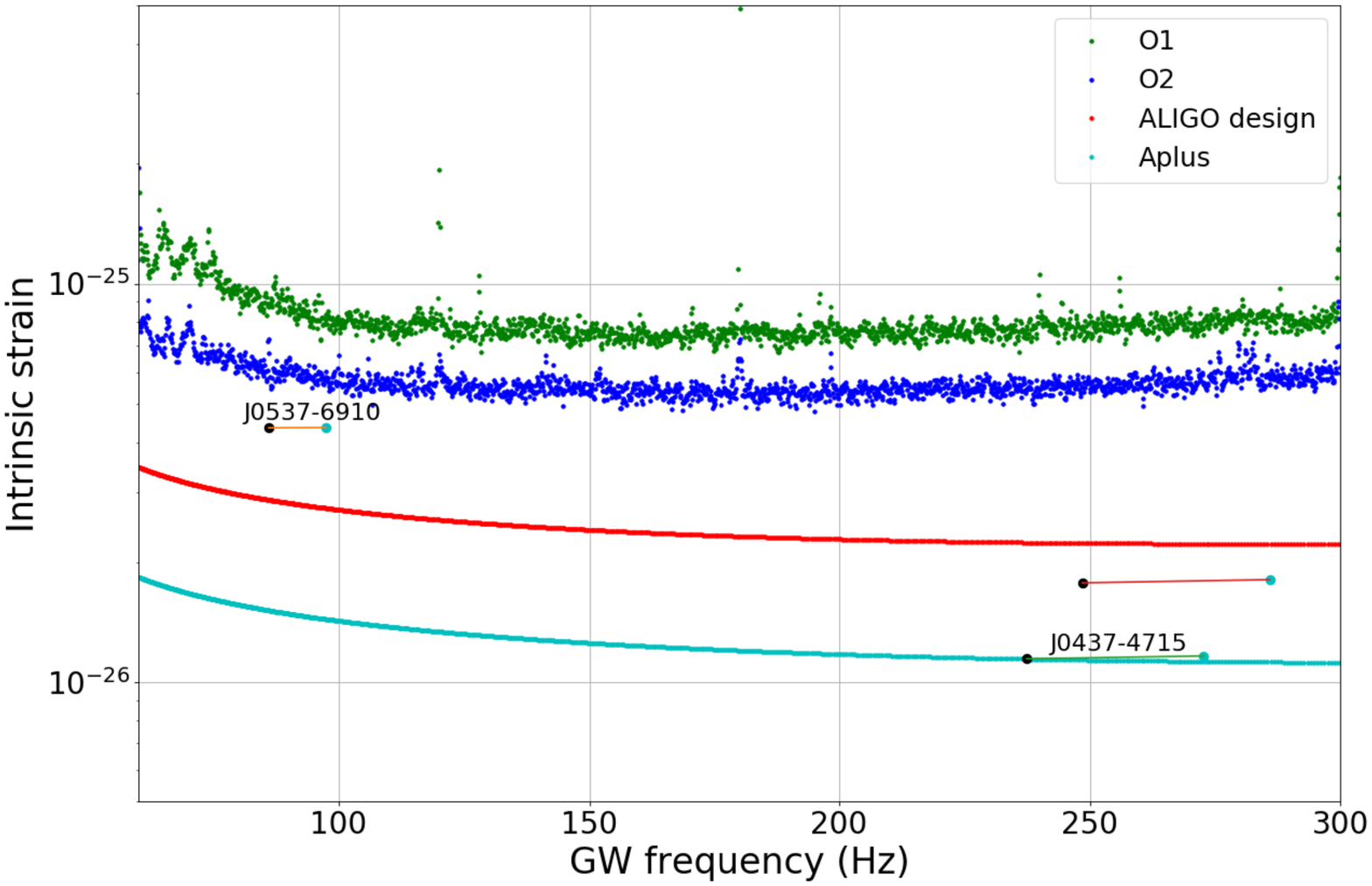}
\caption{
\label{fig2}
Same as the previous figure, for higher frequencies.
There are fewer pulsars here, but the spin-down limits on $r$-mode amplitude
are generally closer to predictions of saturation amplitude.
}
\end{figure*}

We express the sensitivity of each search in terms of upper limits on $h_0$
that can be placed in the absence of a detection.
This is slightly pessimistic---the upper limits are conservative by design and
it is plausible that a somewhat fainter signal could be detected---but it
facilitates comparison with published upper limits from previous searches for
continuous \acp{GW}.
The precise definition of $h_0^\mathrm{UL}$ we use is the same as for instance
in Ref.~\cite{Abbott:2018qee}.
It is a 95\% confidence limit on a population of injected signals with fixed
$h_0$ but varying frequency (within a small band), frequency derivatives, and
angles of inclination and polarization.

The sensitivity of a search of data from a single detector with stationary
noise can be expressed as~\cite{Wette:2011eu}
\begin{equation}
\label{h0Td}
h_0^\mathrm{UL} = \frac{5}{2} \hat\rho \sqrt{ \frac{S_h} {T_d} },
\end{equation}
where $S_h$ is the strain noise \ac{PSD} and $T_d$ is the amount of data.
(In general $T_d$ is less than the integration span $T$ times the number of
interferometers due to maintenance, earthquakes, and so on.)
For multiple detectors or non-stationary noise the \ac{PSD} in
Eq.~(\ref{h0Td}) is replaced by a weighted sum~\cite{Jaranowski:1998qm,
Cutler:2005hc}.
For observations of many days at most sky locations, the sum is very close to
the harmonic mean of noise \acp{PSD}, so we will use the harmonic mean when we
give numbers later.
The statistical factor $\hat\rho$ is iteratively estimated to sufficient
precision using the method of Wette~\cite{Wette:2011eu}, using the template
densities above and assuming that the upper limits are placed on 0.1\,Hz
frequency bands.
(This upper limit band might be chosen differently for different searches, but
its effect on sensitivity is negligible.)
The factor $5\hat\rho/2$ ranges about 33--38 for the searches considered here,
comparable to the factor for directed searches~\cite{Abbott:2018qee} and about
triple the factor for exact timing searches of known
pulsars~\cite{Authors:2019ztc}.
We consider noise \acp{PSD} for \ac{O1}~\cite{H1O1, L1O1}, \ac{O2}~\cite{H1O2,
L1O2}, Advanced LIGO design~\cite{Design}, and the recently funded A+
design~\cite{A+}.
Hence we show four sensitivities:
\ac{O1}, \ac{O2}, and one year integrations at Advanced LIGO and A+ design.

In Figs.~\ref{fig1} and~\ref{fig2} we plot our sensitivity measure vs.\
frequency for the four cases mentioned above, superposed on a set of spin-down
limits for known pulsars from the ATNF catalogue~\cite{Manchester:2004bp}.
Most of the pulsars plotted have already been searched for \acp{GW} at
$f=2\nu$ (and some also at $f=\nu$) in previous LIGO and Virgo papers based on
exact pulsar timing solutions~\cite{Authors:2019ztc}.
Most of the pulsars whose spin-down limits are accessible with existing data
are young and energetic (and sometimes glitchy) like the Crab, and most are
shown in Fig.~\ref{fig1}.
As with known-timing searches, the Crab is the first spin-down limit to become
accessible (already in \ac{O1}), and several more including Vela soon follow.
For later noise curves some middle-aged pulsars (in Fig.~\ref{fig1}) and some
recycled millisecond (in Fig.~\ref{fig2}) pulsars become accessible.
Most notable in Fig.~\ref{fig2} are J0537\textminus6910 and
J0437\textminus4715.
Due to its frequency and proximity to Earth, the latter has
$\alpha_\mathrm{sd}$ of order $10^{-5}$---much lower, and hence more feasible,
than the other accessible pulsars, although $h_0^\mathrm{sd}$ indicates this
pulsar will require at least A+ to detect.

Not all of these pulsars are timed concurrently with LIGO-Virgo observing
runs.
Since the searches proposed here cover broad frequency bands, the uncertainty
in frequency and spin-down parameters is not an issue---unlike the $\nu$ and
$2\nu$ searches.
Our proposed searches do suffer in sensitivity if a pulsar glitches during the
integration, though; and they cannot account for the fluctuating torques
likely in accreting systems.
The glitch issue means that frequent x-ray timing of J0537\textminus6910 will
be important for future \ac{GW} observing runs~\cite{Andersson:2017fow}.

\section{Discussion}

We have shown that searches for continuous \acp{GW} from $r$-modes of known
pulsars can beat the spin-down limits on some pulsars in existing data for
reasonable computational costs.
Although the $r$-mode amplitudes required for detection in such data are
higher than predicted by theory, this work serves as a starting point for
future improvements.
Spin-down limits for many more pulsars will be attainable in the next few
years.

Part of our goal is to point out what theory work could be most important to
help observations.
It is crucial to get the range of mode frequencies and spin-down parameters
right, and helpful to narrow the range down and reduce costs.
Relating frequencies and spin-down parameters more precisely to neutron star
properties will also help measure the latter once a signal is detected.

The most important feature of the search is the $r$-mode frequency range.
Avoided crossings such as that with $t$-modes in the crust could widen the
parameter ranges of some pulsars well beyond what we consider here.
Some pulsars could be undetectable without addressing the avoided crossings
problem.
Updated ranges of the $A$ and $B$ parameters of Eq.~(\ref{fnu}) would also
help in terms of narrowing the parameter space and hence reducing the
computational costs, which will grow to be substantial in coming years.

More estimates of saturation amplitudes would be helpful.
This is a very difficult problem, and essentially has been addressed only by
one approach~\cite{Arras:2002dw}.

It would also help to be sure of the coherence time.
If saturated $r$-modes in equilibrium with other modes occasionally experience
phase jumps, this would render long coherent integrations of \ac{GW} data
problematic.

The coherence time issue leads into future work for data analysis:
Accretion and glitches can also introduce issues which encourage development
of alternatives to the straightforward coherent integrations considered here.
Adaptations of semi-coherent techniques developed for other
searches~\cite{Sun:2017zge, Suvorova:2017dpm, Dergachev:2011pd,
Ashton:2018qth} could be fruitful for $r$-modes from known pulsars too.
For those pulsars with a known inclination angle, upgrading from the
$\mathcal{F}$-statistic to the $\mathcal{G}$-statistic will also improve
sensitivity.

\acknowledgments

We are grateful to the continuous waves search group of the LIGO Scientific
Collaboration, particularly Ian Jones and Karl Wette, for helpful discussions.
This work was supported by NSF grant PHY-1607673.
This paper has been assigned document number LIGO-P1900173.

\bibliography{rmeth}

\end{document}